# Highly durable crack sensor integrated with silicone rubber cantilever for measuring cardiac contractility in culture media

Dong-Su Kim[1a], Yong Whan Choi[2a], Yun-Jin Jeong[1], Jongsung Park[1], Nomin-Erdene Oyunbaatar[1], Eung-Sam Kim[3,5], Mansoo Choi[4] and Dong-Weon Lee[1,5]*

[1]MEMS and Nanotechnology Laboratory, School of Mechanical Systems Engineering, Chonnam National University, Gwangju, 61186, Republic of Korea.

[2]Division of Mechanical Convergence Engineering, College of MICT Convergence Engineering, Silla University, Busan, 46958, Republic of Korea.

[3]Department of Biological Sciences, Chonnam National University, Gwangju, 61186, Republic of Korea

[4]Global Frontier Center for Multiscale Energy Systems, Department of Mechanical and Aerospace Engineering, Seoul National University, Seoul, 08826, Republic of Korea

[5]Center for Next Generation Sensor Research and Development, Chonnam National University, Gwangju, 61186, Republic of Korea.

*Correspondence and requests for materials should be addressed to D.-W. L. (email: mems@jnu.ac.kr).

[a] These authors contributed equally to this work.

## Abstract

We propose a novel cantilever device integrated with a polydimethylsiloxane (PDMS)-encapsulated crack sensor that directly measures the cardiac contractility. The crack sensor was chemically bonded to a PDMS thin layer to form a sandwiched structure which allows to be operated very stably in culture media. The reliability of the proposed crack sensor has improved dramatically

-1-

compared to no encapsulation layer. After evaluating the durability of the crack sensor bonded with the PDMS layer, cardiomyocytes were cultured on the nano-patterned cantilever for real-time measurement of cardiac contractile forces. The highly sensitive crack sensor continuously measured the cardiac contractility without changing its gauge factor for up to 26 days (>5 million heartbeats). In addition, changes in contractile force induced by drugs were monitored using the crack sensor-integrated cantilever. Finally, experimental results were compared with those obtained via conventional electrophysiological methods to verify the feasibility of building a contraction-based drug-toxicity testing system.

Key words: Strain sensor, Crack sensor, Polymeric cantilever, Encapsulation, Cardiomyocytes, Drug screening

## Introduction

Recently, a variety of methods have been reported for assessing drug toxicity by measuring the contractile force of cardiac cells in in vitro environments [1-9]. Methods for the direct measurement of the contractile force include image processing analysis using an optical microscope and laser-based displacement sensors. Alexandre et al. first proposed the use of micro-post arrays to quantitatively measure the contractility of a single cardiac cell using an optical microscope [1]. Although micro-posts made of flexible materials, such as polydimethylsiloxane (PDMS), can measure the contractility of single cells, the deformation of the micro-posts caused by cardiac contractility is very small. Moreover, it is not easy to grow cells on top of micro-posts. Ashutosh et al. proposed a cantilever-type sensor structure to overcome the limitations of micro-posts which measure the contractility of a single cell level [2]. Cardiac cells cultured on top of the surface structures of the cantilever grow into tissue shapes, which induces large mechanical deformation on the polymer cantilever. Therefore, the reliability of the analysis results obtained via this method is high owing to this large deformation. To overcome the limitations of optical microscopy in cantilever displacement measurements, Kim et al. employed a laser vibrometer that could detect displacements of the cantilever smaller than 100 nm. In addition, the materials available for producing cantilevers have been extended from PDMS (kPa) to SU-8 (GPa) [5]. However, analytical methods using optical image processing and laser-based displacement measurements still require additional equipment for high-throughput screening applications and are not easy to employ for rapidly analyzing various drugs simultaneously.

The strain or stress caused by the mechanical deformation of a metal thin film can be converted into an electrical signal, such as resistance, and analyzed using integrated sensors. Recently, metal (Ti/Au) strain sensors were integrated into each of four parallel PDMS cantilevers and the contraction force of cardiac cells was analyzed in real time [8]. Johan et al. have demonstrated that high-throughput screening is possible by arranging the sensor-integrated polymer cantilever arrays in a well plate [9]. However, there is a drawback in that the reliability of the experimental results related to the measurement of the contraction force of the cardiac cells due to a low gauge factor ($< 3$) is not high enough when using metal strain sensors. To confidently measure the contractile forces induced by cardiac cells, high-sensitivity sensors must be integrated onto the flexible polymer. Typical high-sensitivity sensors in the field of microelectromechanical system (MEMS) include silicon nanowires (SiNWs), graphene or carbon nanotube (CNT) composites, crack sensors, etc. [10-21]. Koumela et al. fabricated SiNW sensors with a suspended structure by injecting boron into a SOI wafer and via dry etching. They obtained a maximum gauge factor (GF) of 235 at a boron concentration of $2 \times 10^{20}$–$5 \times 10^{17}$ cm$^{-3}$ [10]. However, because SiNWs formed on hard substrates (such as silicon cantilevers) are hardly deformed by the very small contractile force of cardiac cells, the fabricated sensor must be transferred to flexible substrates. Luo et al. proposed a new method to fabricate a flexible sensor structure via spin-coating on a SiNW formed vertically on a substrate [11]. However, in this method, the final shape of the sensor is that of a simple bar, and it is not easy to integrate such SiNW sensors onto the cantilevers as compared with the conventional metal deposition method.

Crack sensors, which mimic the slit organ of spiders, have been reported to be very easy to fabricate as well as provide a very high sensitivity compared with other sensors [13-15]. Crack sensors employ a thin metal (Cr/Au or Pt) layer as a conductive layer and a polymer substrate layer to induce cracks in the metal layer. D. Kang et al. was the first to propose a crack-based sensor [13]. Cracks were induced by depositing Pt on a poly(urethane acrylate) (PUA) substrate and applying 2% tensile strength to the substrate. The crack-based sensors exhibit surprising sensitivity in air with a GF greater than 2,000 for tensile stress tests. B. Park et al. improved the mechanical sensitivity and signal-to-noise ratio (SNR) of the crack sensors by adjusting the depth of the cracks [14]. T. Lee et al. reported a transparent crack sensor made via ITO deposition on a PET substrate [15]. Y.W. Choi et al. produced a patterned polymer substrate to improve the 2% strain limit, which is a drawback of conventional crack-based sensors. They obtained a GF of $2 \times 10^6$ at a strain of 10% using this surface-patterned polymer [21]. However, because crack-based sensors have a structure in which stress is concentrated between the cracks, sensitivity degradation can be caused owing to the fatigue fracture of the polymer substrate. B. Park et al. solved the problem of sensitivity degradation over long-term use by coating a self-healing polymer on a crack-based sensor [22]. However, when coating with a simple heterogeneous material rather than generating chemical bonds, the bonding surface is not robust and the protective layer can become separated when used for a long time.

In this study, we propose a silicone rubber cantilever integrated with a highly sensitive crack-based sensor to analyze the changes in cardiac contractility induced by very small amounts of drugs in real time. The crack-based sensor made of platinum (Pt) was chemically bonded with a PDMS thin layer by depositing an adhesion layer ($SiO_2$: 2 nm) on the Pt. The plasma bonding process performed at a low vacuum and room temperature not only does not affect the function of the conductive layer, but also avoids contamination of the cantilever surface. In addition, the mechanical durability of the crack sensor was amazingly improved due to the chemical bonding between crack sensor and protection layer. The fabricated crack sensor exhibited a high gauge factor (GF) of $9 \times 10^6$ at a strain of 1% even after the formation of the encapsulation layer. The durability (26 days, >5 million heartbeats) of the sensor was also confirmed in various solutions, such as DI water, culture media. After various basic experiments, the changes in the contractile force of cardiac cells induced by two drugs, namely Verapamil and Quinidine, were evaluated using cantilevers integrated with the PDMS-encapsulated crack sensor. The experimental results were compared with the results obtained via electrophysiological methods, such as using a microelectrode array (MEA). The proposed crack-based cantilever sensor arrays are expected to be applicable in various fields, such as cardiac toxicity tests in the initial stage of the development of new drugs, owing to its excellent stability over long periods of time in an electrolyte solution.

## Results and Discussion

**Concept of highly durable crack sensor working culture media.** The proposed cantilever integrated with the PDMS encapsulated crack sensor consists of a silicone rubber cantilever, a PDMS thin film and a glass body as shown in Fig. 1 (a). The highly sensitive strain sensor based on metal cracks was formed on the silicone rubber cantilever, allowing us to precisely monitor the strain changes caused by the mechanical contraction of cultured cardiac cells on the cantilever. In addition, Au patterns formed on the glass substrate were electrically connected to other Au patterns formed on the silicone rubber via plasma bonding. The use of the glass body with the Au patterns greatly improved the electrical reliability of the fabricated cantilever sensor. Figure 1 (a) also shows the principle of the cantilever sensor used to measure the contractile force of cardiac cells. Figure 1 (b) shows nano-patterns formed on the cantilever surface to align cardiac cells along the length direction of the cantilever. Figure 1 (c) shows equivalent circuit diagrams of two different crack sensors operating in culture media. The electrical pathway of conventional crack sensors consists of a parallel resistor with cracks and liquid, whereas the current flow in the proposed crack sensor only occurs through cracks generated on the cantilever. Crack sensors exposed to culture media behave very unstably like variable resistors. Furthermore, as for the previous PDMS-coating method proposed by S. K. Hong et al., the protection layer is physically combined with the crack sensor via spin-coating, which also affects the long-term durability of the crack sensor [16]. However, the proposed crack sensor was chemically bonded to a very thin PDMS film via plasma bonding and can be used very reliably in various ionic liquids. The use of an intermediate $Cr/SiO_2$ layer on the sensing layer greatly improved the adhesion force between Pt and PDMS.

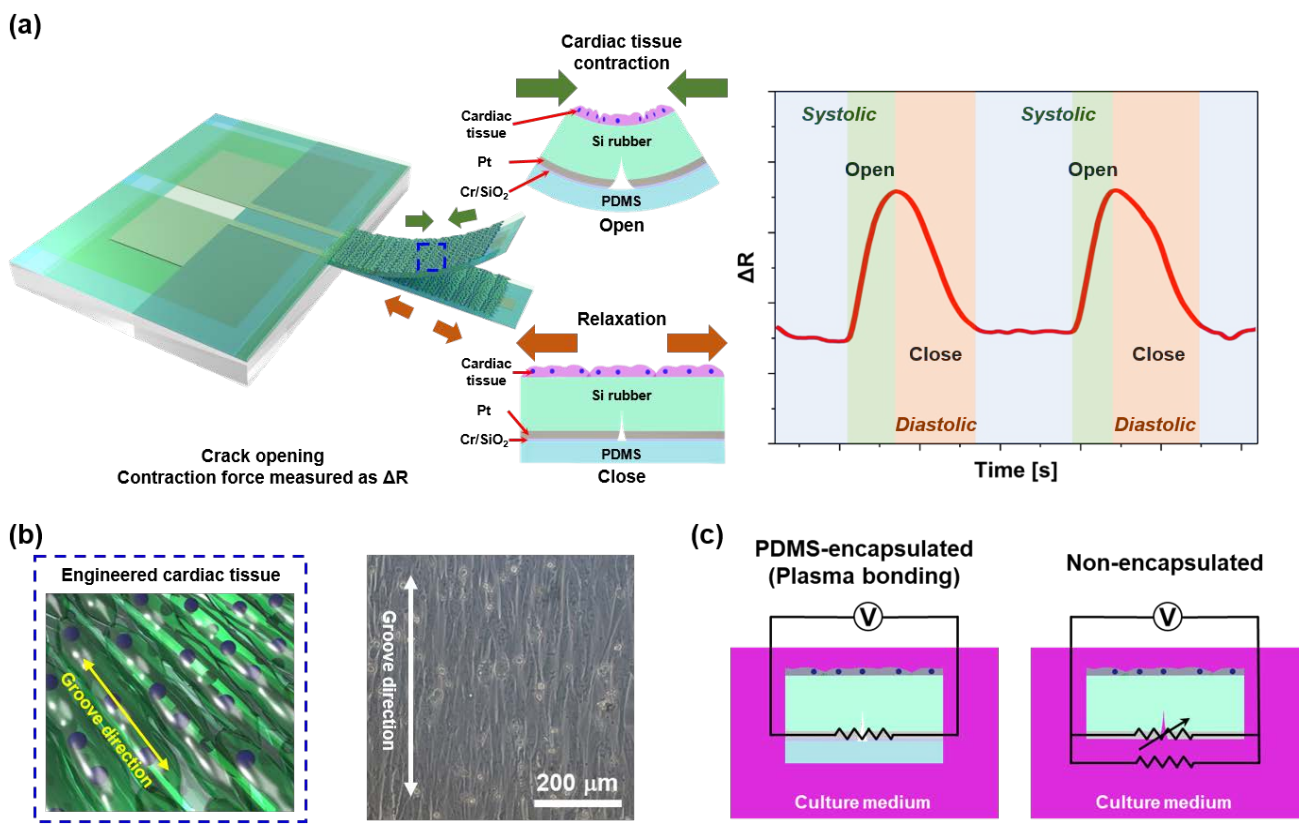

Figure 1. (a) Schematic of a Si rubber cantilever composed of various layers and operation principle of the cantilever sensor integrated with a PDMS-encapsulated crack sensor to measure the contractile force of cardiac cells in liquids. (b) Nano-patterns used to align the cardiac cells on the cantilever surface. (c) Circuit diagrams of two different crack sensors operating in culture media

**Performance of PDMS-encapsulated crack sensor integrated on cantilever.** The sensing mechanisms of crack sensors have been reported several times as shown in Fig. S1 [13-22]. The cracks formed on metal thin films have a shape in which stress is concentrated at the crack ends. When the gap distances of the cracks are increased, electrical resistance is greatly increased owing to the decrease of the tunneling current as well as the number of the electrical contacts of each crack. Optical microscope images were used to monitor the number of contacts of each crack in the fabricated crack sensor. Figure 2 (a) shows an optical microscope image depicting an increase in cracks due to tensile strain, also evidenced by the intensity changes as shown in Fig. 2 (b). An increase in tensile strain in the range of 0–2% was found to proportionally increase the number of cracks. Tensile testing equipment (3342 UTM, Instron Co., Norwood, MA, USA) was employed to measure the electrical conductivity changes of the fabricated crack sensor according to tensile strain changes in the range of 0–0.3% and 0–1%, respectively. The resistance changes on the fabricated crack sensor were measured in real time using a Lab-VIEW-based data acquisition system (PXI-4071, National Instruments Inc., Austin, TX, USA). The standard deviation of the measured value was 0.01 Ω. As shown in Figs. 2 (c, d), the electric resistance of the crack sensor increased 1.7 times for a 0.3% strain and increased approximately 9,000 times for a 1% strain, respectively. The crack sensors used in this test included two types with and without encapsulation layers. It was confirmed to have the same sensitivity in air regardless of the presence of the encapsulation layer. The results of the hysteresis of the crack sensor is shown in Fig. 2 (e). Owing to the inherent characteristics of the polymer material, there was a certain hysteresis value but it was significantly reduced compared with that of the PDMS cantilever. The PDMS-encapsulated crack sensor had a gauge factor of $9 \times 10^6$ at a 1% strain.

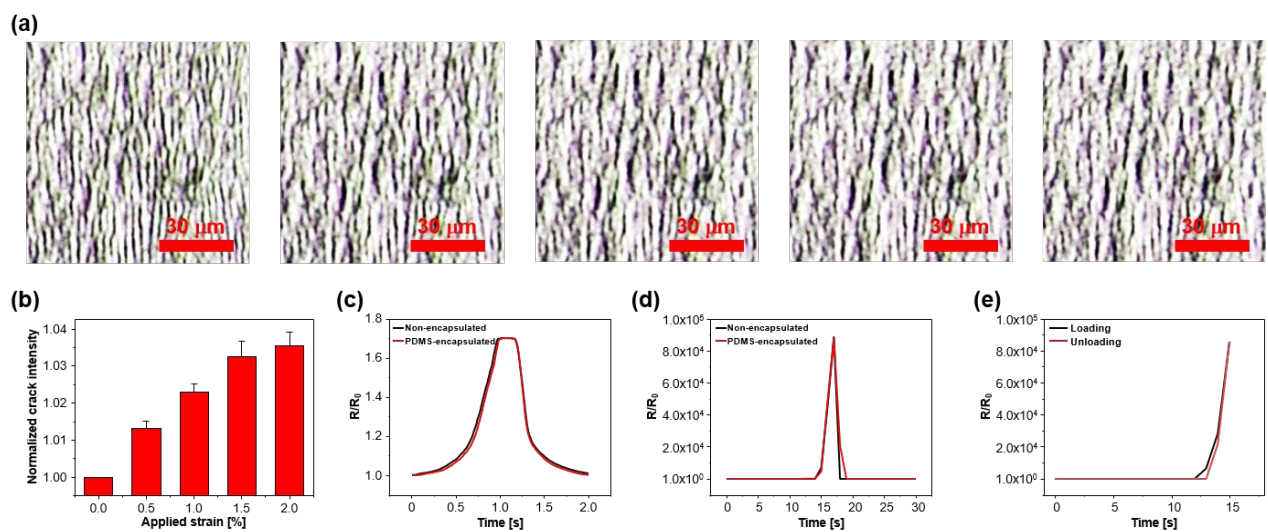

Figure 2. (a) Optical microscope images of the PDMS-encapsulated crack sensor before and after stretching (ε = 0, 0.5, 1, 1.5, 2%). (b) Intensity of cracks according to the applied tensile strain. Resistance changes as a function of strain ranging from (c) 0 to 0.3% and from (d) 0 to 1%, respectively. (e) Hysteresis of the PDMS-encapsulated crack sensor.

Most of the crack sensors reported tend to be irregular in sensitivity depending on the medium between the cracks owing to the exposure of metal parts. However, as for the proposed PDMS-encapsulated crack sensor fabricated through the chemical bonding between sensor and covering layer, we expected that the crack sensor would operate more stably in various environments compared with previously reported crack sensors. To verify the stability of the proposed encapsulation technique, the fabricated crack sensor was exposed to various environments, such as environments with different humidity, temperatures, and conductive liquids. The changes in electrical resistance were evaluated as a function of time. Generally, the metal at the edges of the cracks may be damaged because of the peeling of the thin metallic film from the cantilever surface when used in conditions of repeated tensile strain, temperature changes, or liquid environments. In particular, repeated exposure to harsh environments can result in an intensive stress concentration between cracks owing to different Poisson's ratios and thermal expansion mismatches [20]. The PDMS used as the encapsulation layer

-8-

has high resistance to liquids because of its hydrophobic characteristics. Additionally, because it was chemically bonded to the crack sensor via the $O_2$ plasma treatment, it was robust enough to withstand a tensile strength of 2–5 bar [28]. Experiments were conducted in environments with varying humidity to monitor the changes in resistance of the crack sensor due to changes in the dielectric constant of the medium between the cracks. The PDMS-encapsulated crack sensor showed very stable behavior in the humidity range of 45–95% compared with the non-encapsulated crack sensor. This is probably due to the encapsulation layer preventing vaporized water molecules from penetrating into cracks. On the other hand, the crack sensor without the encapsulation layer allows water molecules to permeate through its cracks, which may cause irregular crack opening and closing. This results in unstable resistance changes of ± 4%, as shown in Figs. 3 (a, d).

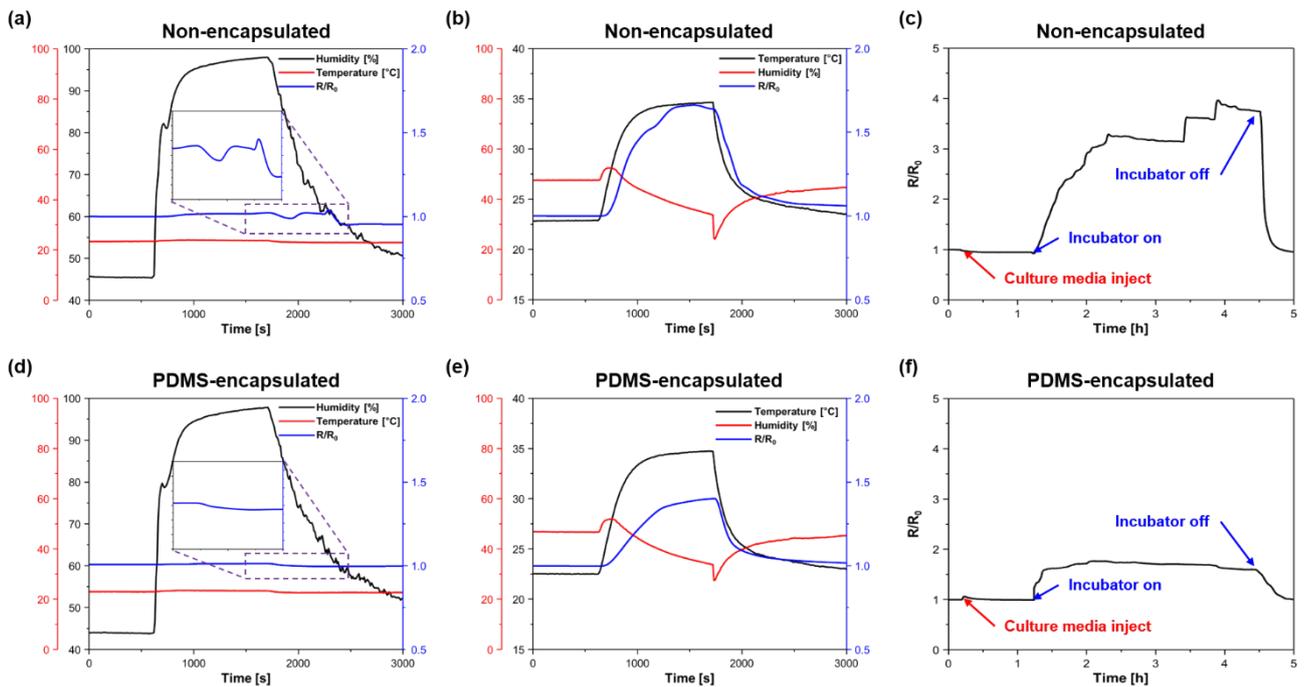

Figure 3. Changes of resistance under different operating conditions for two different cantilevers with and without PDMS encapsulation. Performances of the crack sensors as a function of (a, d) humidity changes (45–95 %) and (b, e) temperature changes. (c, f) Stability of the PDMS-encapsulated crack sensor in a stage-top bioreactor at 37 °C.

The temperature characteristics of the crack sensor according to encapsulation are shown in Figs. 3 (b,e). For the non-encapsulated crack sensor, the resistance change rate was 170% at a maximum temperature of 35 °C. However, the PDMS-encapsulated crack sensor exhibited relatively stable changes in resistance with a change of approximately 120% at a maximum temperature of 35 °C owing to the protection of the encapsulation layer. This can be explained by the fact that the crack sensor was protected from sudden temperature changes by using a polymer with a low thermal conductivity coefficient. In addition, because the proposed crack-based cantilever sensor used for measuring the contractility of cardiac cells is to be used in an incubator environment in which temperature and humidity are maintained, changes in the initial resistance can be neglected as shown in Figs. 3 (c,f). For the non-encapsulated crack sensor, its initial resistance value increased up to approximately four times as the temperature increased from room temperature to 38 °C, as required by the stage-top incubator. Exposing this crack sensor to electrolyte solutions had a significant impact on its reliability because its resistance value changed rapidly even in the culture media with a constant temperature. In contrast, the PDMS-encapsulated crack sensor saturated with a resistance change similar to that of simply increasing the temperature in air. The hydrophobic characteristics and mechanical sealing effect of the encapsulation layer mentioned above seemed to prevent the penetration of the culture media into the cracks even after the operating temperature was increased. Therefore, the crack-based cantilever sensor with the chemically bonded encapsulation layer exhibited improved durability in a variety of environments in terms of humidity, temperature, and culture media.

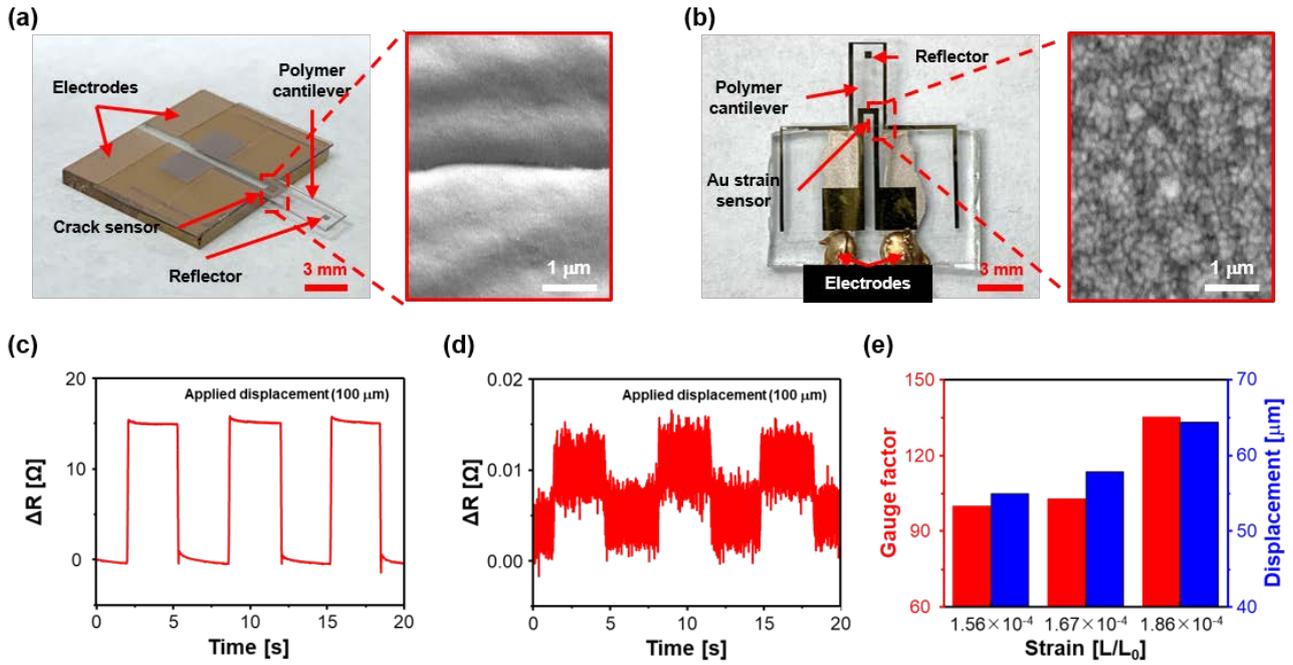

Figure 4. Optical and SEM images of (a) the PDMS-encapsulated crack sensor and (b) an Au strain sensor integrated on a Si rubber cantilever. Resistance changes of (c) the proposed crack sensor and (d) the Au strain sensor as a function of displacement. (e) Gauge factor of the crack sensors depending on strain values.

Figures 4 (a,b) show optical and SEM images of the fabricated crack-based cantilever sensor and a cantilever integrated with the Au/Ti strain sensor used in a previous study. A concentrated load was applied to the free end of both cantilevers using a motorized stage in order to reproduce a contraction and relaxation behavior similar to that of cardiac cells. Figure 4 (c) shows the experimental results for the crack-based cantilever sensor, which exhibited a resistance change of approximately 15 $\Omega$ (GF = 323.9) when a displacement of 100 $\mu$m (strain = $\sim 3 \times 10^{-4}$) was applied to the cantilever with a cycle time of 3 s. On the other hand, Fig. 4 (d) shows the measurement results for the Au/Ti strain sensor; its resistance changes showed a very low value of 16.7 m$\Omega$ (GF = 1.07). We predicted that measurements of cantilever displacement caused by cardiac cells would not clearly distinguishable for assessing the occurrence of the signal owing to environmental noise. The fabricated PDMS-

-11-

encapsulated crack sensor had a 900-times better sensitivity than the metal strain sensor in the same strain range. Figure 4 (e) shows the experimental results of cardiac cell measurements using the PDMS-encapsulated crack sensor in culture media. It exhibited a very high GF of over 130 in the low tensile strain range. The resistance changes as a function of cantilever displacement for two different strange ranges, namely 0–100 μm and 0–1000 μm, are shown in Fig. S2.

**Long-term stability and drug-induced cardiac toxicity applications of the crack sensor.** The durability of crack sensors that are to operate stably over a long period of time is critical for in vitro assays (Fig. S3). In particular, to improve the maturation of cardiac cells, continuous mechanical stimulation due to their contraction and relaxation as well as environmental factors may be applied to the crack sensor (Fig. S4). However, these external stimuli are likely to cause sensitivity degradation by damaging the crack sensor. Figure 5 (a) and (b) show the resistance changes and corresponding displacements of two PDMS-encapsulated crack sensors with different cardiac cell densities. As the density of the cell increases, the displacement of the cantilever increases and the resistance change increases accordingly. As can be seen in the Fig. 5 (a), it can be confirmed that the displacement measurement can be performed stably for 11 days in the culture medium even at a displacement of 5 μm or less. Figure 5 (b) shows the contraction and relaxation characteristics of cardiac cells measured between 17 and 26 days using the cantilever with higher cell density. Figure S5 also shows the changes in resistance over 26 days after culturing the cardiac cells on the proposed crack-based cantilever sensors. The actual displacement of cantilevers was further confirmed by using the laser vibrometer. The beating rate of the cardiac cells was very fast (averaging 4 Hz) at the early stage and then stabilized after 5 days of incubation. Whenever the culture medium was changed every three days after cell seeding, temporary changes in temperature induced a fast beating rate (5 Hz or more) and an unstable status for a certain period. This abnormal state of cardiac cells in contraction was normalized by the stabilization of the temperature of the culture media. The fabricated crack sensor responded very quickly even at a rapid beating frequency of 6 Hz (systolic 82 ms and diastolic 92 ms). The crack-based cantilever sensor showed a greatly stable output for 26 days even for abnormal heartbeats due to changes in the external environment (Fig. S6). Finally, we found that GF changes due to fatigue fracture did not occur, even after 5 million instances of repeated operation.

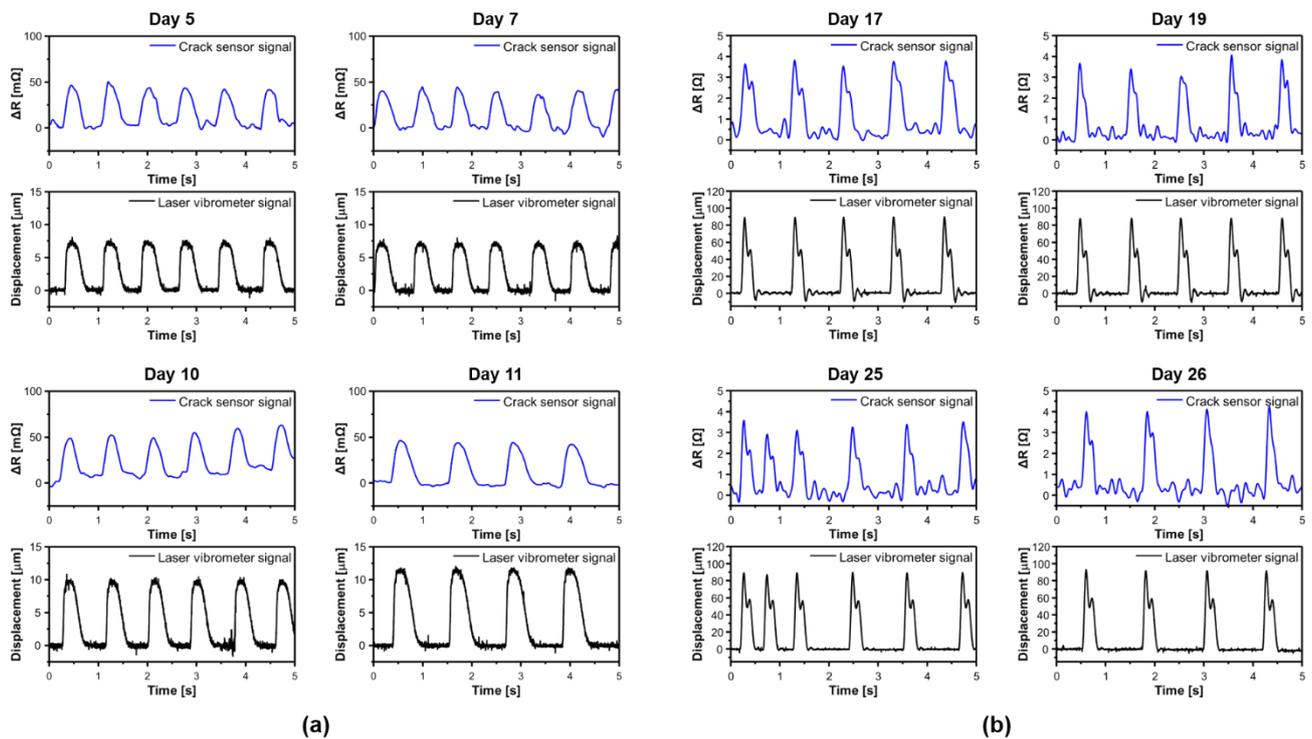

**(a)** **(b)**

Figure 5. Resistance changes and corresponding displacements of the PDMS-encapsulated crack sensor measured over 26 days

After verifying that the cantilever device integrated with the PDMS-encapsulated crack sensor could operate in culture media for a long time, further studies on drugs that affect contractility and beating rate in vitro were conducted based on these preliminary experiments. First, cardiac contractility was measured using the integrated crack sensor as a function of various concentrations of Verapamil, known to be a $Ca^{2+}$ channel blockers. A sudden change in contraction force was observed at a drug concentration of ~1 µM, known to be its IC$_{50}$, as shown in Fig. 6 (a) and Fig. S7 (a). The beating frequency of the cardiac cells also began to show a rapid change at certain concentrations of the drug. Quinidine, known to be a Na$^+$ channel blocker, also induced the same contraction-reducing behaviors at a concentration of ~10 µM, known to be its IC$_{50}$ value. Abnormal beating known as one of the side effects of the drug was also observed with further increases in drug concentration, shown in Fig. 6 (b) and Fig. S7 (b). The drug toxicity screening ability of the crack-based cantilever sensor make it possible to solve the drawbacks of existing optical-based measurement systems and also enables accurate data collection through the simple measurement platform. In addition, by arranging multiple cantilevers integrated with crack sensors in parallel, we expect to be able to simultaneously analyze various drugs with high efficiency.

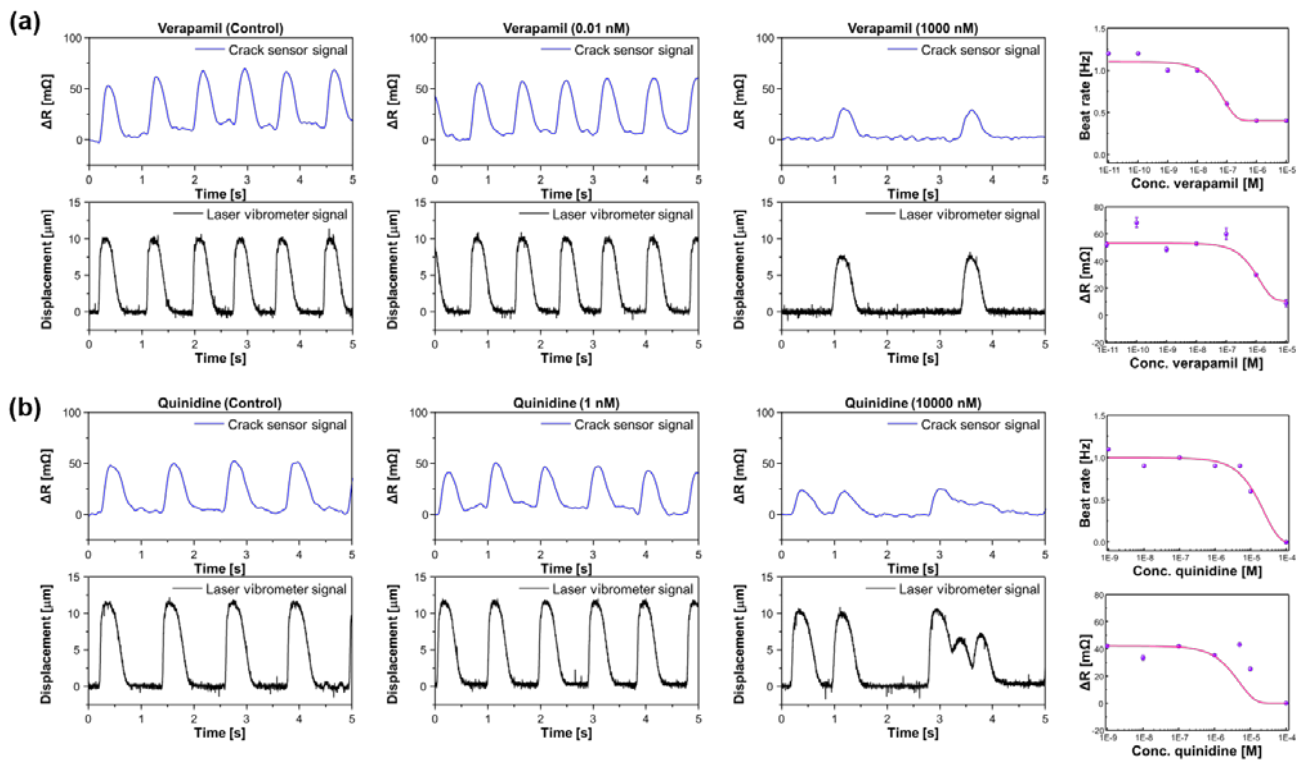

Figure 6. Changes in sensor resistance and cantilever displacement as a function of drug concentration for (a) Verapamil and (b) Quinidine

**Effects of topographic pattern on sarcomere length development and cantilever displacement.**
The cardiac cells were seeded on nano-grooved substrates and grew along the patterns of the substrate, showing a high degree of alignment and a characteristically anisotropic cell structure. The maturation of cardiac cells is dependent on the topographic size of the cultured substrate and has been reported to be the most efficient when the groove size is approximately 800 nm [24]. The sarcomere length of the cardiac cells grown on the nano-patterned cantilever ($2.07 \pm 0.102$ μm) was higher than that of cardiac cells cultured on a flat cantilever ($1.94 \pm 0.077$ μm) as shown in Figs. S8 (a-e). These results indicate that morphological changes in the culture medium enhanced the organization of the cytoskeleton. In addition, the improved organization of the cytoskeleton increased cantilever displacement by a factor of approximately 2.2, which could result in an additional increase in sensitivity in the proposed crack-based cantilever sensor, shown in Figs. S8 (f,g). The proposed PDMS-encapsulated crack sensor was also applied for flexible electronics applications as shown in Fig. S9. The proposed crack sensors chemically bonded to polymer substrates have immense potential for use in various fields related to biosensors.

## Conclusion

The capabilities of the crack-based cantilever sensor proposed in this study were experimentally verified; it can be used for a long time in a solution owing to the PDMS protection layer while also maintaining a high sensitivity. In particular, the chemical bonding of the proposed encapsulation layer, performed to maintain long-term stability in the same environment as the culture medium, showed that it is a very stable encapsulation method that does not affect sensitivity after bonding. A characteristic evaluation of the PDMS-encapsulated sensor in various environments (humidity, temperature, and culture medium) showed the advantage of this crack sensor. In particular, it was possible to stably measure the contractile forces of cardiac cells for approximately four weeks in vitro. The dose-response studies for verapamil and quinidine indicate that the toxicity of drugs can be screened in real time using the proposed crack-based cantilever sensor with a PDMS encapsulation layer. In addition, the alignment of the cardiac tissue on the cantilever's longitudinal direction can induce a larger displacement through the concentration of its contraction force. This can be expected to further increase the gauge factor of the proposed crack sensor, which exhibits exponential behaviors in its resistance changes.

## Methods

**NRVM isolation & Cell culture.** All animal experiments were performed in accordance with protocols approved by the Animal Ethics Committee of Chonnam National University. The NRVM (Neonatal Rat Ventricular Myocytes) was isolated from the heart from a Sprague-Dawley rats within day 1 to 3. The separated ventricles were washed by using $1 \times$ ADS buffer solution (NaCl 120 mM, HEPES 20 mM, NaH$_2$PO$_4$ 8 mM, D-glucose 6 mM, KCl 5 mM, MgSO$_4$ 0.8 mM, DI water 1 L, pH 7.35). Single cardiomyocytes were acquired through enzyme solution (collagenase 0.5 mg / ml, pancreatin 0.6 mg / ml, $1 \times$ ADS buffer solution 50 ml) and pre-plating. In order to effectively coat fibronectin (Corning®), the fabricated cantilever integrated with the PDMS-encapsulated crack sensor was exposed to an oxygen plasma system (FEMTO SCIENCE, 80W, 30 sec) for surface treatment. The acquired cardiomyocytes were then seeded on the crack-based cantilever sensor with a density of 1,000 cells / mm$^2$. Finally, the cardiomyocytes seeded on the cantilever were cultivated at 37 °C in 5% CO$_2$ incubator and the culture medium was replaced every 72 h.

**Immunocytochemical staining.** The cardiomyocytes cultured on the cantilever were fixed in 3.7% formalin solution for 10 min at room temperature (RT) and washed three times using phosphate-buffered saline (PBS Takara). Next, permeabilization was accomplished with 0.1 % Triton-X (Sigma-Aldrich) in PBS for 10 min at RT. To prevent the nonspecific binding of the antibodies, the sample was treated at room temperature for 30 min by using 3 % bovine serum albumin (3 % BSA, Sigma-Aldrich). The primary anti-bodies, monoclonal α-actinin (α-sarcomeric) and vinculin, were diluted 1:100 in 1% BSA and incubated at RT for 1.5 hours. The secondary antibodies were (Alexa-Flour 488 goat anti-mouse IgG conjugate and Alexa-Flour 568 goat anti-rabbit IgG conjugate) diluted 1:200 in the same blocking solution and incubated for 2h at RT. Finally, DAPI (4′,6-Diamidino-2-phenylindole) for nuclei staining was conducted at 37°C for 15min. Immunocytochemical staining analysis was quantitatively performed using an ImageJ software with a normalization from the level of the entire protein [3].

**Process flow of the PDMS-encapsulated crack sensor.** Figure 7 shows a schematic of the silicon rubber cantilever integrated with a PDMS-encapsulated crack sensor. The fabrication process was divided into two parts, one for the polymer part and one for the glass part. A silicon rubber with a relatively high Young's modulus (compared with PDMS) was used for the fabrication of the crack sensor because the Au patterns on the silicon rubber were more stable. As shown in Fig. 7 (a-1), 2 g of silicon rubber compounds (KEG-2000-80A/B, Shin-Etus) were placed on a PUA mold with nano-grooves. A 120-μm-thick feeler gauge (NIKO, Feeler 0.12) was placed on the edge of the PUA substrate, and a polyimide (PI) film was applied and cured by applying a pressure of 4 MPa at 130 °C for 30 min using a thermal press (Fig. 7 (a-2)). Then, a shadow mask was placed on the backside of the Si rubber film patterned with nano-grooves and a Pt thin film with a thickness of 20 nm was deposited using a sputter (Fig. 7 (a-3)). Next, irregular cracks were generated in the Pt thin film by stretching the Si rubber film by approximately 2% using a laboratory-made stretcher, as shown if Fig. 7 (a-4). To increase the bonding strength of the PDMS used as the encapsulation layer, an adhesion layer (Cr/SiO$_2$: 2 nm/2 nm) was deposited on the cracked Pt layer using a thermal evaporator (Fig. 7 (a-5)). The deposited SiO$_2$ layer was chemically bonded to the encapsulation layer (PDMS) via an oxygen-based atmospheric plasma treatment (CUTE-1MPR, Femto Science Inc.), as shown in Fig. 7 (a-6). The shape of the encapsulated crack sensor was precisely defined using a roll-to-plate (TSUKATANI, BFX) and pinnacle die, as shown in Fig. 7 (a-7). The main other process consisted in the fabrication of a glass body in which Au electrodes were formed to enhance the electrical reliability and stability of the crack sensor (Fig. 7 (b)). First, a photoresist was patterned onto a glass wafer and Cr/Au (3 nm/30 nm) was deposited using a thermal evaporator, as shown in Fig. 7 (b-1). After deposition, electrodes were formed on the glass wafer by removing the photoresist using acetone. The glass wafer with Au electrodes was diced into a 9 mm × 12 mm shape using a dicing saw (AM Technology, NDS200), as shown in Fig. 7 (b-2). Finally, the Si rubber cantilever integrated with the crack sensor and the glass body were chemically bonded via an oxygen-based atmospheric plasma treatment (CUTE-1MPR, Femto Science Inc.), as shown in Fig. 7 (c).

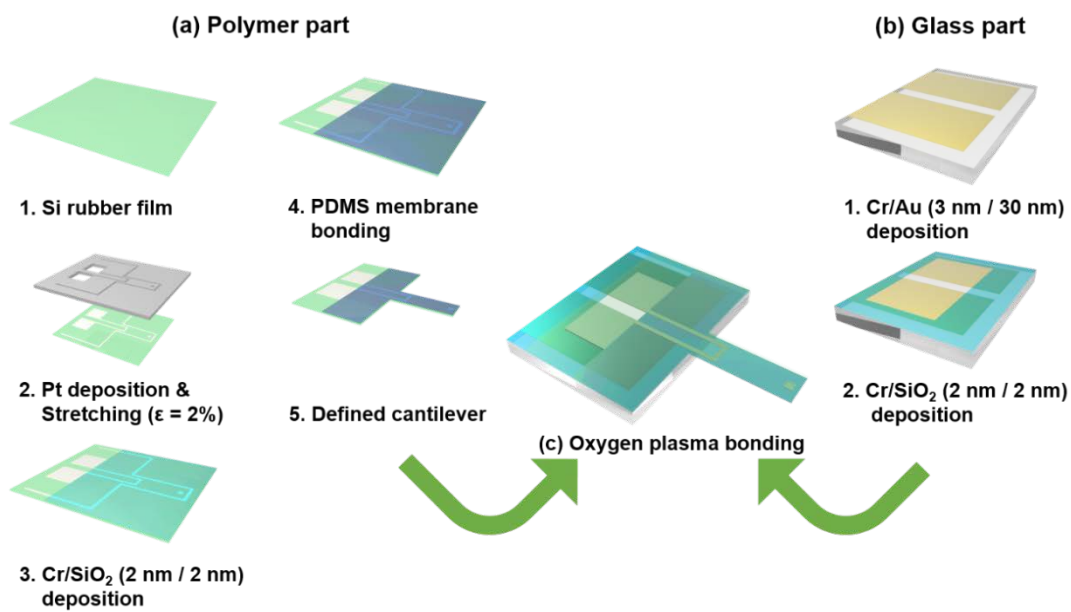

Figure 7. Process flow for the fabrication of the proposed cantilever sensor integrated with a PDMS-encapsulated crack sensor

**Material selection for sensor reliability and cell adhesion.** PDMS (Sylgard 184) and Si rubber (KEG 2000-80) are often used as sensor substrates or structure materials in biotechnology owing to their low

Young's modulus, bio-compatibility and easy processability. In this research, to maximize the sensor yield in the metal process and to stably cultivate the cells, the mechanical behavior of the two materials was compared and the substrate material of the crack sensor was then selected. Their Young's modulus was measured by taking the average slope of the stress–strain curves obtained using a universal tensile tester (Shimadzu, EZ-L) for the stress measurements of each material. The measured Young's moduli were approximately 0.6 MPa for PDMS and 4.5 MPa for Si rubber (Fig. S10). Owing to the low hardness (45 A) and the large thermal expansion coefficient (TEC, $9 \times 10^{-6}$) of PDMS, thermal expansion often occurs on the surface of PDMS during metal deposition, which makes stable metal deposition difficult. In particular, it is difficult to form uniform cracks because of delamination. On the other hand, Si rubber has a relatively high hardness (80 A) and low TCE ($4.2 \times 10^{-6}$) compared with PDMS, so a relatively stable metal deposition can be performed, thereby forming cracks without wrinkles. The designed Si rubber cantilever had a width of 2 mm, a length of 6 mm, and a thickness of 100 μm. It had a spring constant of 35.2 mN/m, which is almost seven times higher than that of PDMS cantilevers with the same dimensions. Cantilevers with a low spring constant are easily deformed by the stress differences within the cardiac cells in the fluid, making it difficult to measure the displacement accurately. The cantilever made of Si rubber had a higher spring constant than that made of PDMS, and it could easily measure the contraction force of cardiac cells owing to the high sensitivity of the PDMS-encapsulated crack sensor.

A previous report indicated that cell maturity is affected by the degree of the surface energy of the material used for cell culture and that there are differences in maturity of up to 30% depending on the material characteristics [23]. Surface energy can be quantified simply through hydrophobic and hydrophilic experiments, which indicate the wettability of water. PDMS and Si rubber basically contain a methyl group and exhibit hydrophobic behavior. In particular, surface modification after $O_2$ plasma treatment is particularly important because it affects cell attachment and ECM (fibronectin) coating very sensitively. Figure S11 (a) shows the water contact angle results before and after the plasma treatment. Both materials show hydrophobicity with a contact angle of 100–110° before the $O_2$ plasma treatment. However, after the $O_2$ plasma treatment, the contact angle decreased to approximately $\leq 11°$, indicating that the surface had been temporally modified as shown in Fig. S11 (b). Figure S11 (c) shows the normalized surface energy of PDMS and Si rubber. Si rubber has a 50% higher surface energy than PDMS and is expected to have benefits for cell adhesion.

After culture cardiac cells on the surfaces of PDMS and Si rubber, maturation and adhesion tests were performed according to the characteristics of the materials. Firstly, immunostaining processes for nuclei, α-actinin, and vinculin were performed to visualize the internal structure of the cardiac cells. Immunostaining images were then analyzed using Image J (NIH, Bethesda, MD, USA) and quantitative data were expressed as mean ± standard deviation (SD). In the immunocytochemistry staining images of a 250 μm × 250 μm area recorded using a confocal microscope (Leica), more nuclei could be observed in the Si rubber substrates than the PDMS ones, which reveals the excellent biocompatibility of Si rubber as cell culture substrates (Figs. 8 (a, b)). Cell adhesion was improved by approximately 300% as shown in Fig. 8 (c). Sarcomere length is an important variable for assessing the contractility and maturity of cardiac cells and is known to be approximately 1.8–2.0 μm in the adult myocardium [24]. The sarcomere length of the cells incubated on the Si rubber cantilever was around 1.94 μm (± 0.077), which is higher than the sarcomere length (1.79 ± 0.011 μm) of the cells cultured on the PDMS cantilever. These results indicate that Si rubber promotes the structural organization of the cytoskeleton, and elongates muscle fiber tissue and sarcomere length (Figs. 8 (d–f)). Within cardiac cells, vinculin is known to bind to the cytoskeleton as an adapter protein for integrin and to affect the contraction and relaxation of cardiac cells [25-27]. In the obtained immunostaining images of vinculin, the expression of this protein on the Si rubber cantilever was 60% higher than that on PDMS (Figs. 8 (g–i)).

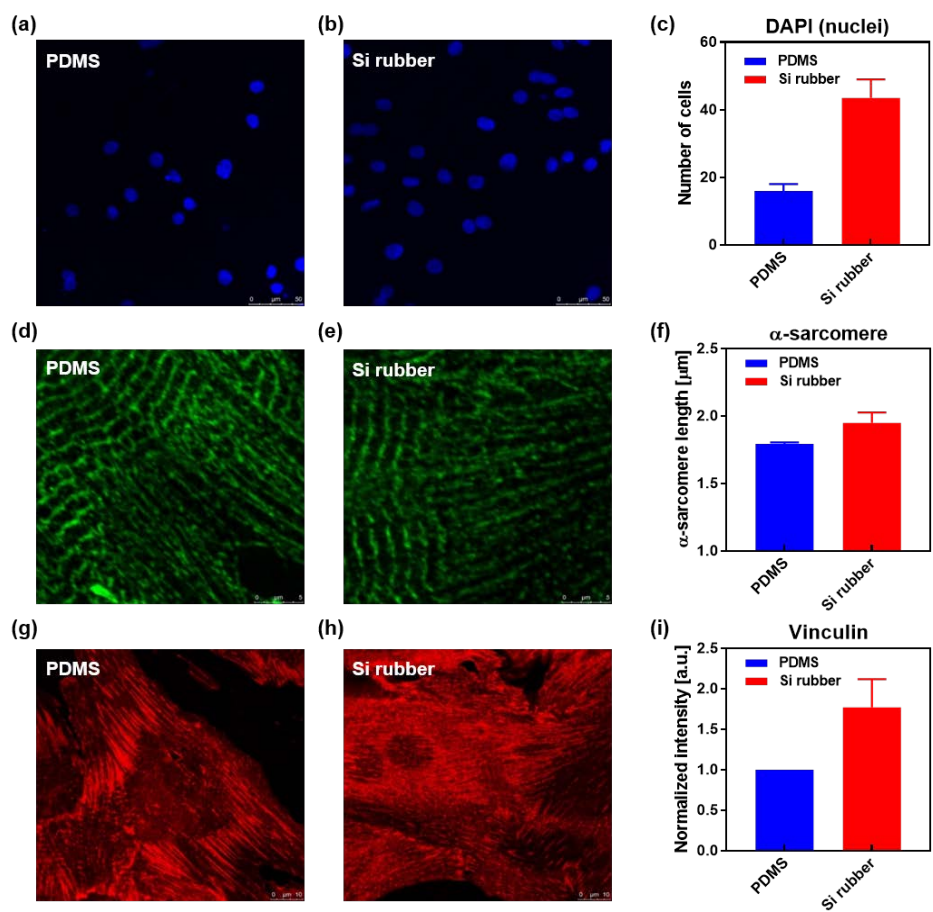

Figure 8. Immunocytochemistry staining images of cardiomyocytes cultured on PDMS and Si rubber substrates. (a–c) DAPI staining, (d–f) α-actinin staining, (g–i) vinculin staining.

## Data availability

All relevant data supporting the findings of this study are available herein and in the Supplementary Information files, or from the corresponding author upon reasonable request.

## Acknowledgement

This study was supported by the National Research Foundation of Korea (NRF) grant funded by the Korea government (MSIT) (No.2017R1E1A1A01074550)

## Author contributions

D. Kim designed the research, discussed the results with E. Kim, M Choi and D. Lee, and also contributed to writing the manuscript with D. Lee. Y Choi prepared and characterized the PDMS-encapsulated crack sensor, N. Oyunbaatar contributed in cardiac cell experiment, Y. Jeong and J. Park assisted the cantilever fabrication.

## Additional information

**Supplementary Information** accompanies this paper at xxxxxx.

**Competing interests:** The authors declare no competing interests.